\documentclass{ifacconf}
\usepackage[nolist,printonlyused]{acronym}
\usepackage{amsfonts}
\usepackage{graphicx}  
\usepackage{amssymb}
\usepackage{mathtools}
\usepackage{natbib}        % required for bibliography
\usepackage{xcolor}
\newcommand\topic[1]{\textcolor{purple}{#1}}

\newcommand\body[1]{#1}
\renewcommand\body[1]{}

\newcommand\conclude[1]{\textcolor{olive}{#1}}

\renewcommand\topic[1]{#1}
\renewcommand\conclude[1]{#1}

\newcommand\subdataset[1]{\dataset{}_{#1}}
\newcommand\wptset[1]{N_{#1}}
\newcommand\traj[1]{\xi^{#1}}
\newcommand\autotraj[1]{{\xi_a}^{#1}}
\newcommand\trajspace{\Xi}

\newcommand\state[1]{\textbf{x}_{#1}}
\newcommand\stategt[2]{\textbf{x}^{\text{GT}, #2}_{#1}}
\newcommand\ctrl[1]{\textbf{u}_{#1}}

\newcommand\numstates{n}
\newcommand\numctrls{r}
\newcommand\numdatapts{m}

\newcommand\horizon{T}
\newcommand\natset[2]{\mathcal{N}^{#1}_{#2}}

\newcommand\dynamics[1]{f_{#1}}

\newcommand\nstates[1]{\mathcal{X}^{#1}}

\newcommand\dataset{\mathcal{D}}

\newcommand\nset{naturalistic behavior set}
\newcommand\nsets{naturalistic behavior sets}

\newcommand\NSET{Naturalistic Behavior Set}

\begin{document}
%===============================================================================
\begin{frontmatter}

    \title{Act~Natural!~Projecting~Autonomous~System Trajectories~Into~Naturalistic~Behavior~Sets%
    % Behaviors Into Learned Naturalistic Sets
    % Naturalistic Projection into Human Behavior Sets 
    % Identifying \NSET{}s and Projecting Behaviors Into Them
    \thanksref{footnoteinfo}} 
    % Title, preferably not more than 10 words.
    
    \thanks[footnoteinfo]{This work was supported by the National Science Foundation under Grant No. 2211548. (e-mail: \{ hamzah, adam.thorpe, dfk \}@utexas. edu).}
    
    \author{Hamzah I. Khan\quad} 
    \author{Adam J. Thorpe\quad} 
    \author{David Fridovich-Keil}
    
    \address{University of Texas at Austin, 
       Austin, TX 78712 USA}

    \begin{abstract} % Abstract of not more than 250 words.
Autonomous agents operating around human actors must consider how their behaviors might affect those humans, even when not directly interacting with them.
To this end, it is often beneficial
% they must choose their actions 
to be predictable and appear naturalistic.
Existing methods to address this problem use human actor intent modeling or imitation learning techniques, but these approaches rarely capture all possible motivations for human behavior or require significant amounts of data.

In contrast, we propose a technique for modeling naturalistic behavior as a set of convex hulls computed over a relatively small dataset of human behavior.
% Our primary contribution is a naturalistic projection technique which first generates \nset{}s from datasets of human behavior.
Given this set, we design an optimization-based filter which projects
% The method then incorporates an optimization problem that enables projecting 
arbitrary trajectories into it to 
make them more naturalistic for autonomous agents to execute while also satisfying dynamics constraints.
% make them more naturalistic and retain dynamic feasibility.
We demonstrate our methods on real-world human driving data from the inD intersection dataset \citep{inD2020naturalisticdataset}.

% \todo{Write abstract at the end.
% These instructions give you guidelines for preparing papers for IFAC
% technical meetings. Please use this document as a template to prepare
% your manuscript. For submission guidelines, follow instructions on
% paper submission system as well as the event website.} \\
% \todo{Write abstract at the end.
% These instructions give you guidelines for preparing papers for IFAC
% technical meetings. Please use this document as a template to prepare
% your manuscript. For submission guidelines, follow instructions on
% paper submission system as well as the event website.}
\end{abstract}


\end{frontmatter}
%===============================================================================

\section{Introduction} \label{sec:intro}
% Argument
% \david{Be careful about over-using the word naturalistic.} \\
% \david{Use autonomous system trajectory rather than autonomous trajectory.} \\
\topic{
% To produce safe and comfortable behavior around human actors in interactive settings, 
In interactive settings, human actors require a degree of predictability from autonomous agents
% Autonomous actors that operate around humans  must act predictably 
to ensure the safe and comfortable operation of all interacting actors.}
Failing to do so can cause problems: for example, autonomous cars can often behave more cautiously than humans expect, 
% leading
% Autonomous cars have been shown to be overly cautious, leading
% to collisions related to unexpected behaviors
leading nearby human drivers to react in unexpected ways and, potentially, cause collisions \citep{teoh2017google}.
The requirement of acting like other actors is even encoded in some traffic laws\footnote{Texas Transportation Code §545.363} in which drivers must follow the ``flow of traffic'' regarding their speeds on highways.
As acting in ways that stand out unnecessarily can lead to safety and comfort concerns, autonomous vehicles must be able to understand how humans naturalistically behave.
% However, 
% Thus, safety and comfort in many driving environments requires 
% exhibiting naturalistic behavior, and that achieving this goal necessitates 
% understanding naturalistic human behavior.
% which capture potentially significant aspects of the world that an autonomous vehicle may not capture (i.e. potholes can be a safety concern).
% Consider the case of
% 
% Practical example.
% 
However, naturalistic behavior tends to be opaque and thus difficult to model mathematically, and there is a need for methods that can highlight what naturalistic behavior looks like from observations of human behavior.
Existing solutions \citep{bajcsy2021analyzing,sadigh2016infogatheringonhumans} tend to either model a limited subset of possible influences in human decision-making or do not produce naturalistic behaviors without significant data \citep{kuefler2017imitating}.
\conclude{These deficiencies present a need for improved naturalistic behavior generation in autonomous prediction and downstream planning tasks.}
% Human preferences tend to be opaque and difficult to model mathematically, but their  behaviors capture (potentially significant) aspects of the world that an autonomous vehicle may not capture (i.e. potholes can be a safety concern).
% \conclude{}

\begin{figure}[!t]
\includegraphics[width=\columnwidth]{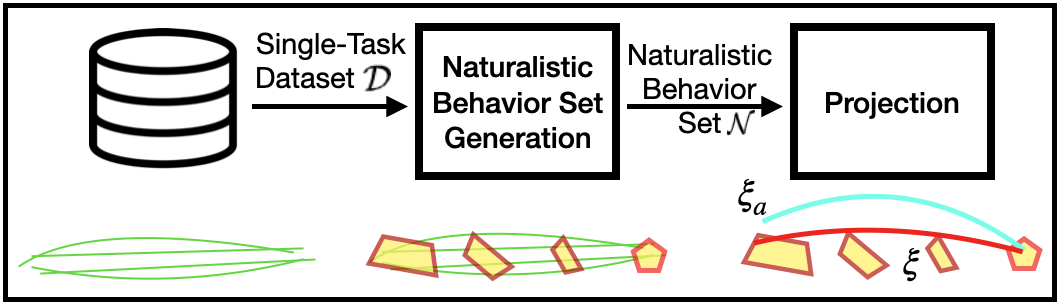}
\includegraphics[width=\columnwidth, height=6.5cm]{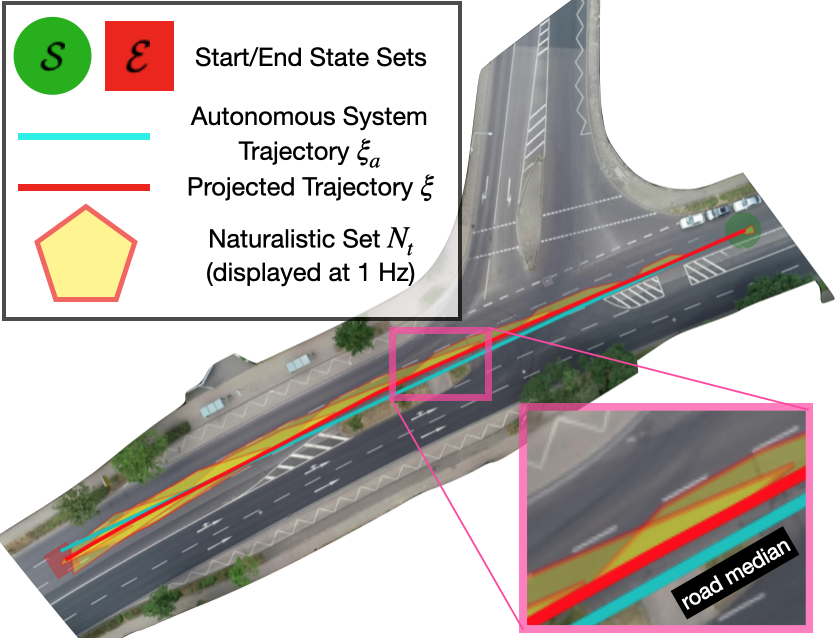}
\caption{\label{fig:hullproject-r1}
(top). Given a single-task dataset $\dataset{}$, our method first generates a \nset{} by computing time-indexed convex hulls.
Then, we project arbitrary trajectories into this set to make the behaviors more naturalistic.
% To make a \nset{}, we filter a naturalistic dataset $\newdataset{}$ to create a single-task dataset $\dataset{}$. We then generate the \nset{} $\natset{}{}$ and can project autonomous trajectories into it.
% Then, we can project autonomously planned trajectories into it.
(bottom). We compute the \nset{} $\natset{}{}$ 
% over a hull state containing 2D position 
using trajectories in the second lane of the upper road beginning in the green circle and ending in the red square. We then project a trajectory into $\natset{}{}$.
% \david{Alter naturalistic set colors to differ from trajectories.}
}
\end{figure}

% Our method...
\topic{We present a data-driven method for identifying the set of naturalistic human behaviors using convex hulls and an optimization-based method for projecting autonomously planned trajectories into it.}
Specifically, given a set of observed human driving trajectories, we compute a sequence of convex hulls around the states at each time instant. 
\conclude{With this representation in hand, we project trajectories (i.e. from autonomous planners) into the \nset{} to enforce naturalistic driving constraints.}

% \begin{figure}[!t]

% \caption{The top figure depicts the process of constructing a \nset{}. We form it from time-indexed convex position hulls for a small dataset of trajectories with horizon $\horizon=4$. The bottom figure depicts the projection of a reference trajectory $\refstate{1:4}$ into the \nset{}. \todo{Make these computerized images.}}
% \end{figure}

% Related work
% \subsubsection{Related Work.}
% 1. Personalized vs. legible driving work

% \adam{I would recommend condensing a lot of this information down. It's a bit long as-is, so I would say if there's a way to smush this down I would do that. I like the first paragraph, and the first sentence of the second paragraph does a good job of categorizing the related work. So I would suggest smushing P2 and P3.} \\
% \hmzh{one more look through literature to make sure story ties together well} \\
\topic{Existing work establishes that autonomous planners must consider the effect of their behavior on human actors \citep{sadigh2016planning}.} \cite{dragan2013legibility}
% dragan2013legiblemotion
propose the concept of legible motion from which an observer can easily infer intent or strategy through actor motion.
This concept differs from predictability, which involves easily inferred motion without necessarily understanding intent.
\cite{sadigh2016planning} notes that the actions of an autonomous vehicle influence human 
drivers on the road, potentially unintentionally. 
\conclude{Such a conclusion implies a need for autonomous vehicles to model human behavior in prediction and planning, as otherwise safe but unexpected actions might cause unsafe reactions by human actors.}

\topic{Existing methods tend to approach the problem of accounting for human behavior from one of two directions: inferring intent from human behavior models and imitating behavior based on observed data.}
Human behavior models use approaches ranging from explicitly modeling and inferring specific aspects of 
human internal states like rationality \citep{bobu2018learning} and target states \citep{sadigh2016infogatheringonhumans} to data-driven predictive human behavior 
modeling that infers human actors' beliefs of other agents' goals
\citep{bajcsy2021analyzing,fridovich2019confidenceaware}, or influencing human 
actors' (potentially incorrect) internal models of autonomous agents' motion \citep{tian2023dynamicsofhumanlearning}.
% 
% However, there are no mathematical models
However, these approaches tend to fall short because human preference and behavior is 
naturally opaque and difficult to model mathematically, resulting in multiple unmodeled aspects of the world which may be significant, like how road conditions may influence safety and comfort tolerances. 
% Identifying specific aspects of the human internal state that influence intent may be possible \citep{sadigh2016infogatheringonhumans,bajcsy2021analyzing,fridovich2019confidenceaware,tian2023dynamicsofhumanlearning}, but these approaches focus on inferring specific aspects of human models like target states and rationality.
% However, they neglect to model other relevant factors in human decision-making, such as how road conditions may influence safety and comfort tolerances.
% and 
% the full suite of possible influences on human behavior is too large to practically model. 
\conclude{Our approach differs by seeking to frame naturalistic behavior planning as a projection into naturalistic constraints from human behavior data, without otherwise inferring the internal state of human actors.}

% 2. Inverse reinforcement learning and behavior cloning
% \todo{Find better/more recent citations.}
\topic{The second set of approaches to this problem is imitation learning, and these involve motion planning by mimicking provided human behavior data through behavior cloning or inverse reinforcement learning methods.}
% In behavior cloning, models are trained on human data in a supervised manner.
% However, the performance of 
Behavior cloning methods train supervised models on human data but tend to suffer from compounding errors and distribution shift 
\citep{bagnell2015invitation,kuefler2017imitating}.
Inverse reinforcement learning, which attempts to infer a cost function describing agent objectives, generalizes better than behavior cloning and produces naturalistic human behavior in microscopic settings (i.e., using the local context) 
% \todo{Use wheeler/kochenderfer paper for this wording}
but requires significantly more data \citep{kuefler2017imitating}.
Additionally, most inverse reinforcement learning approaches assume very specific reward structures \citep{song2018imitationmultiagent,huang2021driving}.
% Thus, we find in the prior work that the opaqueness of human behavior makes inverse reinforcement learning similarly challenging. 
\conclude{For abnormal scenarios in which we have limited data due to dynamic conditions (i.e., bad weather, rare road scenarios), even imitation learning techniques are infeasible for modeling human behavior due to insufficient data.}

% \subsubsection{Contribution.}
% The contribution...
% \david{it is not a big deal at all, but just as an example of a place where you have substantial redundancy, look at the first two sentences of this paragraph.}
\topic{Our main innovation is a simple and efficient method for identifying convex constraints that capture naturalistic human driving behaviors. We then use our representation to formulate and solve a trajectory projection problem that enforces naturalistic behavior, similarity to an original trajectory, and dynamic feasibility.}
% Some more contributions.
We make two key contributions: 1) constructing a convex hull around naturalistic driving data, and 2) an optimization-based method for projecting autonomous solutions onto the learned sets.
% How are these novel? What's new here? Could contrast, or point out gaps in existing work.
\conclude{Through numerical results on the {inD} real-world scenario driving dataset \citep{inD2020naturalisticdataset}, we demonstrate the ability of our \nset{} generation 
% \adam{I don't know what this means. Is this analysis? I don't think there is any analysis here to reveal underlying patterns.}
method to capture patterns, which may not be explicitly modeled, in human behavior data.
% that are not explicitly modeled, and to 
We additionally adjust trajectories generated from other sources to capture such patterns through projection.}

\section{Problem Statement} \label{sec:problem-statement}
% In this work, we seek a method of representing a \nset{} and identifying it from human behavior data.
% Then, we propose a projection to efficiently map a trajectory onto the \nset{}.

Consider a discrete-time dynamical system,
\begin{equation}
\state{t+1} = \dynamics{}(\state{t}, \ctrl{t}),
\label{eq:dynamics}
\end{equation}
where $\state{t} \in \nstates{} \subseteq \mathbb{R}^{\numstates{}}$ is the state at time $t$ and $\ctrl{t} \in \mathcal{U} \subseteq \mathbb{R}^{\numctrls{}}$ is the control input.
The system
% is discretized at period $\Delta t$, and it 
evolves from an initial condition $\state{0}$ over a finite time horizon $t = 0, \ldots, \horizon$, where $\horizon \in \mathbb{N}$.
% We denote a trajectory of states over the full horizon as $\state{1:\horizon} = (\state{1}, \state{2}, \ldots, \state{\horizon})$, and the control trajectory over the time horizon as $\ctrl{1:\horizon} = (\ctrl{1}, \ctrl{2}, \ldots, \ctrl{\horizon})$.

We assume that the dynamics $\dynamics{}$ in \eqref{eq:dynamics} are known.
% , but that the control inputs $\ctrl{t}$ are given by a human operator.
% \david{just say ``we know $f$ in (1), and have a dataset of recorded human-driven trajectories (2)''}
% Consequently, the future state evolution of the system is unknown.
We additionally presume access to historical human behavior data consisting of state trajectories of the form 
\begin{equation}
\label{eq:trajectory-form}
\traj{i} = [ \state{0}^{i\intercal} ~ \state{1}^{i\intercal} ~ \cdots ~ \state{\horizon}^{i\intercal} ]^\intercal \in \trajspace{} \subseteq \mathbb{R}^{\numstates{}(\horizon{} + 1)}.
\end{equation}
We further expect this data to describe ``well-behaved'' human behavior.
For example, well-behaved human behavior should generally be safe, as humans naturally avoid unsafe behavior.
% \hmzh{Is this too casual/vague? I'd appreciate suggestions on how to word this.}
% As a concrete example, most human drivers are able to drive safely

% As humans perform consistently well on common tasks like driving (i.e., ), we assume the historical data \hmzh{continue}
% As historical data of human behavior will perform reasonable 
% contains reasonable examples of human behavior (i.e., human drivers might slightly exceed the speed limit, ).}
% For example, we assume that in the driving application, human operators typically seek to operate safely,
% While we formulate $\traj{i}$ as a set in \eqref{eq:trajectory-form}, we may equivalently treat it as a vector $\traj{i} \in \mathbb{R}^{\numstates{}(\horizon{} + 1)}$ formed by stacking each element in chronological order.

% While we do not have access to the control inputs corresponding to \eqref{eq:trajectory-form}, we 

% \begin{equation}
% \label{eq:times}
% t \in \{1, 2, \ldots, \horizon\} \equiv \mathbb{T}. 
% \end{equation}
% % $t \in \{ 1, 2, \ldots, \horizon \} $.
% The state evolves according to the dynamics
% \begin{equation}
% \state{t+1} = \dynamics{}(\state{t}, \ctrl{t})
% \label{eq:dynamics}
% \end{equation}
% We denote a trajectory of states over the full horizon as $\state{1:\horizon} = (\state{1}, \state{2}, \ldots, \state{\horizon})$, and the control trajectory over the time horizon as $\ctrl{1:\horizon} = (\ctrl{1}, \ctrl{2}, \ldots, \ctrl{\horizon})$.

\subsection{Estimating the \NSET{}}
\label{ssec:formulation-estimating-nset}

Our goal is to compute a set-based representation of the state trajectories that captures the naturalistic behavior of the human operator performing a given task, e.g. a driver making a left turn at an intersection or navigating a particular path from point A to point B. 
% \adam{[[Give an interpretation or definition of task.]]}
In other words, given a dataset $\dataset{}$ which consists of $m$ state trajectories (as in \eqref{eq:trajectory-form}) that 
% \david{Remove safely and put a full sentence explaining your point on this below (2).}
perform a particular task,
\begin{equation}
    \label{eq:dataset}
    \dataset{} = \{ \traj{i} 
    % = \{ \state{1}^{i}, \state{2}^{i}, \ldots, \state{\horizon}^{i} \}
    \}_{i=1}^{\numdatapts{}},
\end{equation}
we seek to estimate the \nset{} $\natset{}{}$.
We formulate $\natset{}{}$ as a collection of sets 
\begin{equation}
\label{eq:full-nat-set}
\natset{}{} = \{ \wptset{0}, \wptset{1}, \ldots, \wptset{\horizon} \}    
\end{equation}
indexed by time, such that each $\wptset{t}$ is a subset of the state space $\mathcal{X}$ and $\natset{}{}$ forms a tube over the full time horizon $\horizon$.

The problem of representing sets of feasible trajectories is fundamentally one of forward reachability \citep{bansal2017reachability}.
% [cite something, maybe from my advisor C. Tomlin. There’s a good tutorial paper from CDC 2017]. 
% The primary challenge comes from the fact that the human inputs are unknown.
% Robust approaches  for estimating the future evolution of a system exist, and can generally be described via uncertainty propagation. 
Robust approaches that seek a strict over-approximation, while safe, tend to be overly conservative, and rapidly lose value for long-term prediction or estimation \citep{rober2023reachability}. 
Bounded interval techniques suffer from the same issues \citep{ramdani2011reachable}.
Our approach can be viewed as a purely data-driven approximation of the naturalistic set, in line with some sample-based forward reachability methods (e.g. \citealp{lew2022simple, thorpe2022data}).

\subsection{Projection Into the \NSET{}}

Because autonomous system trajectories do not demonstrate naturalistic behavior, the resulting behaviors may be unpredictable, and therefore unsafe, when operating around humans \citep{teoh2017google}.
% We seek to modify existing autonomous trajectories to 
% Next, we seek to adjust an existing autonomous trajectory to make it more naturalistic. 
% \adam{[[Problem statement of making autonomous planners behave more like humans.]]}
% Autonomous planners don't demonstrate naturalistic behavior.
Thus, we seek to augment an autonomously generated trajectory (i.e., from an arbitrary autonomous planner) to make the trajectory behave more naturally. 
% More specifically,  can be accomplished via a projection of the planned trajectory into the \nset{}.
Specifically, given a trajectory $\autotraj{}$ computed from an autonomous planner, we seek to project $\autotraj{}$ into the learned \nset{} $\natset{}{}$.
Further, we seek a projection that satisfies dynamic feasibility with respect to \eqref{eq:dynamics} while remaining as similar to the original trajectory as possible.

\section{Method} \label{sec:method}
\newcommand\trajlength{H}
\newcommand\numhullstates{n_c}
\newcommand\hullstate[1]{y_{#1}}

% \adam{Do you say this anywhere else? Specifically in the contributions or abstract? Otherwise, I would remove this name as a proper noun.}
We propose a naturalistic projection technique that identifies a \nset{} from human behavior data and then projects trajectories into the representation
% produced by \ac{NaBSetId} method 
in a dynamically consistent manner.

\subsection{\NSET{} Identification}
\label{ssec:method-naturalistic-set}
\newcommand\trajhorizon[1]{H^{#1}}
\newcommand\newdataset{\tilde{\dataset{}}}
\newcommand\hullsubstatefn[2]{\hullstate{}(\state{#1}^{#2})}

\newcommand{\startset}{\mathcal{S}}
\newcommand{\finalset}{\mathcal{E}}

% We first identify the \nset{} $\natset{}{}$ given data $\dataset{}$.
% We then elaborate on design choices and challenges related to the selection of the naturalistic dataset, the computation of convex constraints, and the computational complexity. 
% \adam{[[Using a dataset $\dataset{}$, we first seek to learn or identify the naturalistic set $\natset{}{}$.]]}

We presume access to a naturalistic dataset $\dataset{} = \{ \traj{i} \}_{i=1}^{\numdatapts{}}$ as in \eqref{eq:dataset} consisting of $\numdatapts{}$  trajectories of length $\horizon$.
Using $\dataset{}$, we define time-indexed datasets $\{ \subdataset{t} \}$ at times $t = 0, 1, \ldots, \horizon$ by gathering all states at time $t$ across trajectories in $\dataset{}$. 
% We then select a hull state of size $\numhullstates{}$ produced through a transformation (i.e. selector matrix, principal component analysis, etc.) of the state $\state{t}$, $\hullstate{t} = \hullsubstatefn{t}{}$. 
For each time $t = 0, 1, \ldots, \horizon$, we then seek to learn $\wptset{t} \in \natset{}{}$ using data 
\begin{equation}
    \label{eq:dataset-slice}
    \subdataset{t} = \{ \state{t}^{0}, \state{t}^{1}, \ldots, \state{t}^{m} \}.
\end{equation}

\subsubsection{Representing the \NSET{}.}
\label{ssec:filtering-data}

% becomes dependency on location and behavior type that allows our \nset{} representation to be used to analyze local patterns of behavior.

% Next, we note that dataset $\dataset{}$ may contain a wide variety of naturalistic behaviors across different scenarios for an application.
% As an example within the driving application, consider attempting to represent all possible behaviors around intersections in a given city.
% Depending on the location of the intersection, the specific maneuver being performed, local road conditions, and other factors, naturalistic behavior may vary significantly.
% The variety of initial conditions and possible actions make any overly broad representation likely to wash out local patterns and variations and thus reduce the usefulness of a \nset{}.
% % too general and difficult to interpret in a useful manner.
% % local patterns and variations get washed out by statistical methods 
% % Thus, our method constructs a useful representation of naturalistic behavior by first specifying which behavior we seek to represent.

% \hmzh{Does this capture everything Adam suggested?}
% \adam{[[Filtering the data. Outliers. Safety. Other considerations for human data.]]}
We represent a \nset{} as a tube of convex hulls.
This choice provides a number of practical benefits, including simplicity, flexibility, and data efficiency.
First, taking a convex hull produces an equivalent set of linear inequality constraints and ensures that every data point is represented in it.
% While outliers are known to inflate convex hulls, we note that outlier rejection techniques like RANSAC \citep{fischler1981ransac} exist.
Second, convex hulls do not assume a particular distribution of the underlying data, which is critical to capturing the wide variety of human behavior.
% While this operation may result in an over-approximation of the \nset{}, we note that similar over-approximations are used by forward reachability techniques \todo{cite?} to similarly simplify assumptions on the structure of complex data or challenging dynamics.
Third, using convex hulls avoids requiring a large amount of data.
Thus, our method provides the benefit of working on smaller datasets, as compared to more data-intensive learning methods.
% \david{This is confusing (explain microscopic). Remove?}
% As such, we focus on ensuring that our method captures microscopic behaviors in $\dataset{}$ and we propose that this benefit makes our method useful as a fine-tuning tool for adjustments of more complicated motion plans in these microscopic settings.
We discuss additional considerations regarding our choice of convex hulls, including those related to non-convexity, in Section \ref{sec:limitations}.

% After selecting the hull state, we compute time-indexed convex hulls as a means of simply and efficiently representing the \nset{} at each time.
% We note that each facet of each hull can be represented as a set of half-space intersections $\{(G_t, h_t) ~|~ G_t \hullstate{t} \leq h_t \}$.

% \adam{[[Halfspace constraints.]]}

\subsubsection{Efficiently Computing Complex Hulls.}
% Computational Considerations in 
% \subsubsection{Computing Convex Hulls.}
In general, the worst-case computational complexity of producing a convex hull from $\numdatapts{}$ points in $\mathbb{R}^\numstates{}$ is $O(\numdatapts{}^{\lfloor \numstates{}/2\rfloor})$ \citep{barber1996quickhull}.
% $d = |\tilde{\dataset{}}|$ and a hull state of size $\numhullstates{}$ over horizon $\horizon$.
Our method computes a convex hull at each time step, which requires $\horizon$ computations.
Thus, the overall computational complexity of \nset{} generation is $O(\horizon \numdatapts{}^{\lfloor \numstates{}/2\rfloor})$.
% Each computation of a convex hull has a worst-case complexity of $O(\numdatapts{}^{\lfloor \numhullstates{}/2} \rfloor)$ \citep{barber1996quickhull}.

We note the number of facets on the boundary of a convex hull can rise exponentially with $\numstates{}$.
% \hmzh{Too many half-spaces?}
Each facet corresponds to a half-space, so introducing additional facets similarly increases the number of constraints required to represent a convex polytope as a half-space intersection (i.e., linear inequality constraints).
% This can cause computational issues as adding half-spaces can similar increase the number of inequality constraints.
% \todo{We need to discuss the half-space intersection constraints before this.}

% \todo{make y a different letter, like M, z, P}

% \hmzh{Kushagra - don't start with instead}
To address this problem, we consider a transformation (e.g. using a selector matrix or principal component analysis) of the state $\state{t}$.  
We call the transformed state a ``hull state'' in $\mathbb{R}^{\numhullstates{}}$, where $\numhullstates{} < \numstates{}$, produced via $\hullstate{t} = \hullsubstatefn{t}{}$. 
In other words, for all $t = 0, 1, \ldots, \horizon$, we seek to learn $\wptset{t} \in \natset{}{}$ using data 
\begin{equation}
    \label{eq:dataset-slice-hull}
    \subdataset{t} = \{ \hullsubstatefn{t}{0}, \hullsubstatefn{t}{1}, \ldots, \hullsubstatefn{t}{m} \}.
\end{equation} 

In general, having few hull states (i.e., small $\numhullstates{}$) results in a computationally efficient algorithm.
However, if more hull states are needed, polynomial-time approximation algorithms for computing convex hulls exist \citep{Sartipizadeh2016ComputingTA,balestriero2022deephull}.
% can eliminate the dimensionality term and provide a polynomial time method.
% Thus, the overall computational complexity of \nset{} generation is $O(\horizon d^{\lfloor \numhullstates{}/2} \rfloor)$.

% \adam{[[Short paragraph discussing the computational efficiency of the hull method.]]}

We 
% elect to
represent $\wptset{t}$ via the 
% $\numhullstates{}$-dimensional 
convex hull of hull states $\hullstate{t}^{i}$,
\begin{equation}
    \label{eq:convex-hull}
    \wptset{t} = H(\subdataset{t}) = \text{ConvexHull}(\{ \hullstate{t}^{i} \}_{i=1}^{\numdatapts{}}).
\end{equation}
The \nset{} $\natset{}{}$ can then be constructed as a collection of 
convex 
hulls indexed in time, as in \eqref{eq:full-nat-set}. 
% Note that in general, we can also consider a convex hull over a transformation of the state $g(\state{t})$, e.g. in order to...

% \hmzh{Is this paragraph out of order?}
\subsubsection{Forming Convex Constraints.}
% In computing the convex hull, we treat the set of hull states at a given time as a convex set, noting that such a treatment may lead to over-approximating the true naturalistic set.
% \todo{mention that this is common in forward reachability analysis}
% We then construct convex hulls over the hull states at each time $t$, $\{\hullstate{t}^{\text{GT}, (i)}\}_{i=1}^{|\newdataset{}|}$.
% We assume that we can construct an over-approximation of the true naturalistic set via a convex hull computed over the hull state.
Convex sets can be well-approximated as convex polytopes, which can be equivalently represented via half-space intersections.
% , we write $\wptset{t} = (G_t, h_t)$.
Thus, the condition $\hullstate{t} \in \wptset{t}$ can be equivalently written as the (linear) half-space intersection inequality constraint
\begin{equation}
\label{eq:half-space-intersection}
G_t \hullstate{t} \leq h_t.
\end{equation}
Expressing the \nset{} in this form allows us to utilize \eqref{eq:half-space-intersection} as a constraint within an optimization problem.
% , as we discuss further in Section \ref{ssec:method-projection}.

\subsection{Projection Into the \NSET{}}
\label{ssec:method-projection}
Next, we seek to make a given autonomous system trajectory $\autotraj{}$ behave more naturally while retaining dynamic feasibility.
This can be interpreted as projecting the trajectory into the \nset{}.
% \david{total nit: I suggest using $\xi$ to represent the original trajectory (instead of $\autotraj{}$) and then $\hat \xi$ to be the approximated (projected) version. I think this would be more aligned with conventions in many areas, but it’s also a matter of taste.}
We thus seek to identify a set of controls $\ctrl{0}, \ldots, \ctrl{\horizon-1}$ that generates a naturalistic trajectory $\traj{}$ similar to $\autotraj{}$ subject to the dynamic constraints $\dynamics{}$. Let $\state{init}$ be the initial condition at time $t=0$ of $\xi_{a}$.
We define the projection
% \david{Remind the reader how $\traj{}$ depends on $\state{t}$}
\begin{subequations}
\label{eq:projection-opt}
\begin{align}
\min_{\ctrl{0}, \ldots, \ctrl{\horizon-1}} \quad & \text{d}(\autotraj{}, \traj{}) \label{eq:projection-opt-objective} \\
\textrm{s.t.} \quad & \traj{} = [ \state{0}^\intercal ~ \state{1}^\intercal ~ \cdots ~ \state{\horizon}^\intercal ]^\intercal \label{eq:projection-opt-traj} \\
% \traj{} = \{ \state{0}, \state{1}, \ldots, \state{\horizon} \} 
              & \state{t+1} = \dynamics{}(\state{t}, \ctrl{t}), \quad t = 0, 1, \ldots, \horizon \label{eq:projection-opt-dynamics} \\
              & \state{0} = \state{init} \\
              & y(\state{t}) \in \wptset{t}, \quad t = 0, 1, \ldots, \horizon,
              \label{eq:projection-opt-nset}
\end{align}
\end{subequations}
where $d : \trajspace{} \times \trajspace{} \to \mathbb{R}$ in \eqref{eq:projection-opt-objective} is a distance metric for trajectories, e.g. the Euclidean distance
% \eqref{eq:projection-opt-traj} formalizes the definition of $\traj{}$ from \eqref{eq:trajectory-form}. 
or 
% the Hausdorff distance for 
a trajectory similarity metric like the one proposed by \cite{chen2011clustering}.
% , which incentivizes $\traj{}$ to be similar to $\autotraj{}$.
The constraints in \eqref{eq:projection-opt-dynamics} and \eqref{eq:projection-opt-nset} enforce dynamic feasibility and the learned naturalistic behavior constraints, which can be modeled as linear inequalities as in \eqref{eq:half-space-intersection}.
% \begin{remark}
% \label{remark:additional-constraints}
In addition, we note that we can easily augment \eqref{eq:projection-opt} to enforce additional constraints such as control limits or safety restrictions.
% Here, we make note of a benefit of our approach.
% As our problem is formulated as an optimization, a user with a definition of a safety set can introduce it as an additional constraint of the optimization problem \ref{eq:projection-opt}.
% This provides our method with versatility that many other methods lack.
% \end{remark}

% and easily modify the result of the projection to satisfy both sets of constraints.

% Next, we discuss some practical details regarding solving \eqref{eq:projection-opt}.
% \adam{[[Continue here with more information about optimization problem. ]]}
% \hmzh{Should this next paragraph be removed?}
% \david{$\trajhorizon{}$ is overloaded.}
% \begin{remark}
% \label{rmk:extended-traj}
% If a trajectory has a longer horizon than the \nset{}, i.e. $\trajhorizon{} > \horizon$, we modify \eqref{eq:projection-opt-nset} to ensure naturalistic behavior until time $\horizon$.
% Dynamic feasibility constraints remain applied over the entire trajectory.
% \end{remark}
The projection in \eqref{eq:projection-opt} may be non-convex if the dynamics $\dynamics{}$ are nonlinear.
Nevertheless, a variety of well-studied techniques exist to identify local minimizers of non-convex problems like \eqref{eq:projection-opt}.
% , which is generally nonconvex.
We refer the reader to \cite{nocedal1999numerical} for further details.

\section{Experiments} \label{sec:experiments}
\newcommand\movingvehspace{\mathcal{V}}
\newcommand\mass{M}

\topic{We demonstrate our naturalistic projection technique on real-world human driving data from the inD dataset \citep{inD2020naturalisticdataset}.}
% \conclude{We highlight that our method can identify naturalistic aspects of human behavior and adjust existing trajectories to be more naturalistic.}

\subsection{The inD Dataset}
\label{ssec:inD-dataset}

% \david{connecting the state space of (12) with (10). This can be a minor modification or addition to the last sentence of this paragraph (with the ref to Sastry ch 9).} \\

% \david{again H overloaded in (14)}

The inD datset 
% \adam{more description needed}
\citep{inD2020naturalisticdataset} records and labels naturalistic traffic data for vehicles, bicyclists, and pedestrians at four German intersections using a drone camera positioned above each intersection.
% \david{not germane though it is interesting}
% As compared to driving studies or ground-level sensors, this approach effectively records traffic actors without the measurement affecting naturalistic driving behavior \citep{inD2020naturalisticdataset}.
% \hmzh{Explain why this benefits our method.}
% While some of these actors do not move (i.e. parked cars), most enter the field of view, travel across the intersection, and exit.

Each actor $i$'s trajectory is annotated at time $t$ with state
\begin{equation}
\label{eq:inD-state}
\stategt{t}{i} = [\vec{p}_t^{i\intercal} ~ \vec{v}_t^{i\intercal} ~ \vec{a}_t^{i\intercal} ~ \theta_t ]^\intercal,
\end{equation}
containing planar position $\vec{p}_t^i$ tracking the center of the actor, planar velocity $\vec{v}_t^i$, planar acceleration $\vec{a}_t^i$, and heading $\theta_t^i$.
Trajectory $\xi^i$ for actor $i$ is constructed as in \eqref{eq:trajectory-form} and states are sampled at $25$ frames per second, with actor $i$ being visible and recorded from the first frame in which actor $i$ is visible, at $t=0$, 
% being the first frame in which  and visible horizon 
until the last frame in which it is visible, at $t=\trajhorizon{i}$.

\subsection{Identifying a Single-Task Dataset $\dataset{}$}
\label{ssec:experiments-dataset}
The inD dataset contains naturalistic trajectories of actors performing a variety of tasks. 
% not separate naturalistic trajectories into sets of tasks, 
We  generate a single-task dataset $\dataset{}$ by defining a heuristic indicator function.
% as described in Section \ref{ssec:filtering-data}.
Let $\movingvehspace{}$ contain the indices of all moving vehicles. We formalize a filtering heuristic
% we specify a task of interest for each experiment based on a start set $\startset{}$ and an end set $\finalset{}$.
% We consider every vehicle beginning within the start polygon and ending in the end polygon to be performing the same maneuver.
\begin{equation}
\label{eq:exp-heuristic}
h(\xi^i; \startset{}, \finalset{}) = i \in \movingvehspace{} ~\wedge~ \state{0}^i \in \startset{} ~\wedge~ \state{\trajhorizon{i}}^i \in \finalset{}.
\end{equation}
To ensure nontrivial behavior, the first term of \eqref{eq:exp-heuristic} considers only moving vehicles.
The second and third terms further filter the naturalistic trajectories under the assumption that every actor moving from a start polygon $\startset{}$ to an end polygon $\finalset{}$ performs the same task.
% of moving vehicles based on a selected task.
% As the moving vehicles from the dataset perform a variety of actions (i.e., lane changes, turns, etc.), 
% we consider every vehicle moving from a start polygon $\startset{}$ to an end polygon $\finalset{}$ to be performing the same task.
We specify different polygons $\startset{}, \finalset{}$ for each subsequent experiment.
% \david{Remove from here to end of section.}
% We note that this approach is appropriate for the driving application because we can treat a task as beginning in one lane and ending in another on either the same road or a different one (e.g., a left turn, proceeding straight in a given lane, etc.).
% We generate $\dataset{}$ 
% % from $\newdataset{}$
% using binary indicator heuristic \eqref{eq:exp-heuristic} to identify trajectories executing the task of interest.
% % Given the heuristic defined by \eqref{eq:exp-heuristic}, 
% % \eqref{eq:dataset-from-heuristic}.

\subsection{Generating the \NSET{}}
\label{ssec:exp-nset-generation}
% \hmzh{Is describing the unicycle dynamics transformation necessary?}
% We model each moving vehicle actor with planar unicycle dynamics discretized at period $\Delta t$.
% Given naturalistic data provided in the format described by \eqref{eq:inD-state}, we produce 
% To produce a \nset{}, we must first define a hull state 
Our method requires naturalistic data over which we can compute convex hulls, so we define the hull state using the information available in \eqref{eq:inD-state}.
% , from which we would be forming the \nset{}.
We first select dynamics $\dynamics{}$ by modeling each moving vehicle actor as a point with mass $\mass$ evolving according to planar double-integrator dynamics
\begin{equation}
\label{eq:exp-planar-di-dynamics}
\state{t+1} = \left[ \begin{array}{c}
    p_{x,t+1} \\
    v_{x,t+1} \\
    p_{y,t+1} \\
    v_{y,t+1} \\
\end{array} \right]
= \left[ \begin{array}{c}
    p_{x,t} + \Delta t v_{x,t} \\
    v_{x,t} + \Delta t F_{x,t} / \mass \\
    p_{y,t} + \Delta t v_{y,t} \\
    v_{y,t} + \Delta t F_{y,t} / \mass \\
\end{array} \right],
\end{equation}
where $\ctrl{t} = [F_{x, t} ~ F_{y, t}]^\intercal$ are forces applied to the point mass.
As \eqref{eq:exp-planar-di-dynamics} constitutes a linear equation in $\state{t}$ and $\ctrl{t}$, we denote the dynamics as $\state{t+1} = A\state{t} + B\ctrl{t}$ for brevity.
In practice, many systems of interest are differentially flat and admit a representation of state and control in which dynamics are linear \citep[Ch. 9]{sastry2013nonlinear}.
% \todo{Confirm that Chapter 9 is the correct one.}
% We specifically note that the state could be 

Next, we define the hull state
\begin{equation}
\label{eq:exp-hullstate}
\hullstate{t} = \hullsubstatefn{t}{} = \left[ \begin{array}{cccc}
    1 & 0 & 0 & 0  \\
    0 & 0 & 1 & 0 
\end{array} \right]  \state{t}
\end{equation}
by extracting the two-dimensional position from state $\state{t}$ of the dynamics in \eqref{eq:exp-planar-di-dynamics}.
As the full \nset{} $\natset{}{}$ captures sets of positions over time, we neglect higher order kinematics, although they can be included in principle at the expense of additional computation.
We note that including higher order kinematics is only possible given appropriate naturalistic data.
% , as noted previously when discussing the hull state.
Since \eqref{eq:inD-state} includes velocity and acceleration, we could use planar quadruple-integrator dynamics with jerk controls for experiments on this dataset.
% we neglect higher order kinematics though they can be included.
Using \eqref{eq:exp-hullstate}, we generate datasets $\{ 
\subdataset{t} \}_{t=1}^\horizon$ as described by \eqref{eq:dataset-slice-hull}.
As a convex hull can only be generated from a dataset with at least $\numhullstates{}+1$ points, we specify $\horizon$ to be the maximum time satisfying $|\subdataset{t}| \geq \numhullstates{}+1$.
Next, we generate the naturalistic set at each time, $\wptset{t}$, by computing the convex hull with the Quickhull algorithm \citep{barber1996quickhull} as described by \eqref{eq:convex-hull}.
Finally, we build the full \nset{} $\natset{}{}$ as in \eqref{eq:full-nat-set}.

\newcommand\autohorizon{\trajhorizon{}_a}
\subsection{Framing the Projection Problem}
\label{ssec:exp-projection}
For the subsequent experiments, we frame the projection of $\autotraj{}$ into $\natset{}{}$ as an alteration of \eqref{eq:projection-opt},
\begin{subequations}
\label{eq:projection-opt-exp}
\begin{align}
\min_{\ctrl{0},\ldots,\ctrl{\autohorizon{}-1}} ~ & \| \autotraj{} - \traj{} \|_2^2 \label{eq:projection-opt-objective-exp} \\
\textrm{s.t. } & \traj{} = [ \state{0}^\intercal ~ \state{1}^\intercal ~ \cdots ~ \state{\autohorizon{}}^\intercal ]^\intercal \label{eq:projection-opt-traj-exp} \\
% \traj{} = \{ \state{0}, \state{1}, \ldots, \state{\autohorizon{}} \} \label{eq:projection-opt-traj-exp} \\
              & \state{t+1} = A\state{t} + B\ctrl{t} ~ \forall ~ t \in \{0, 1, \ldots, \autohorizon{}\}{} \label{eq:projection-opt-dynamics-exp}\\
              & G_t \hullstate{t} \leq h_t ~ \forall ~ t \in \{0, 1, \ldots, \horizon\}, \label{eq:projection-opt-nset-exp}
\end{align}
\end{subequations}
where $\autotraj{}$ has horizon $\autohorizon{}$ and $\horizon = |\natset{}{}|$.
% Next, we elaborate on each line of the projection.
Equation \eqref{eq:projection-opt-objective-exp} defines the similarity objective $d(\cdot,\cdot)$ as a Euclidean distance.
Equation \eqref{eq:projection-opt-dynamics-exp} enforces (linear) planar double-integrator dynamics over the entire trajectory horizon $\autohorizon{}$ as described by \eqref{eq:exp-planar-di-dynamics}. 
Note that we adjust the projection to account for the case where $\horizon \neq \autohorizon{}$.
% , as discussed in Remark \ref{rmk:extended-traj}.
Equation \eqref{eq:projection-opt-nset-exp} describes naturalistic behavior constraints as linear inequalities, as described by \eqref{eq:half-space-intersection}.

We note that the Euclidean distance $d(\cdot, \cdot)$ is convex. 
Moreover, both constraints \eqref{eq:projection-opt-dynamics-exp} and \eqref{eq:projection-opt-nset-exp} are linear, indicating that \eqref{eq:projection-opt-exp} is a convex optimization problem. 
For this reason, we solve \eqref{eq:projection-opt-exp} with an efficient convex optimization library, CVXPY \citep{diamond2016cvxpy}.
% \hmzh{}
If $\traj{i} \in \dataset{}$ is dynamically feasible, then projection is guaranteed to find a dynamic feasibile trajectory $\traj{} \in \natset{}{}$.
% satisfying .
% Projection is guaranteed to have a solution for dynamic feasibility 

% $\exists \traj{} \in \natset{}{}$ s.t. $\state{t+1} = f(\state{t}, \ctrl{t})$ given that $\traj{i} \in \dataset{}$ is D.F.]]}

% \adam{[[Projection is guaranteed to have a solution for dynamic feasibility. $\exists \traj{} \in \natset{}{}$ s.t. $\state{t+1} = f(\state{t}, \ctrl{t})$ given that $\traj{i} \in \dataset{}$ is D.F.]]}

\subsection{Curved Road}
\label{ssec:exp-recording1}
Fig. \ref{fig:hullproject-r1} depicts two curved roads separated by a median running through a T-intersection.
We define $\dataset{}$ as including all moving vehicles beginning and ending in the second lane of the upper road, where $\startset{}$ is given by the green circle and $\finalset{}$ by the red square.
Filtering based on these criteria results in 39 trajectories.
We expect the \nset{} to be influenced by the curve of the road.
The set generated from $\dataset{}$ is shown in Fig. \ref{fig:hullproject-r1}.

\subsubsection{Analysis.}
The naturalistic sets $\{\wptset{t}\}$ begin compact but lengthen along the lane over time, indicating that vehicles drive at different speeds along this lane.
At the most curved point in the lane, the \nset{} covers the outside portion of the lane but not the inside.
This observation suggests that drivers naturally hug the outside of a curved lane.

\subsubsection{Projection.}
Fig. \ref{fig:hullproject-r1} also depicts $\autotraj{}$, a constant-velocity trajectory moving straight through the second lane.
We call this trajectory non-naturalistic because it is not within the \nset{}.
In particular, we note that it gets abnormally close to the median, which could negatively impact safety or comfort.
As expected, applying a naturalistic projection to $\autotraj{}$
% As shown in Fig. \ref{fig:hullproject-r1}, we naturalistically project this trajectory into the \nset{}, 
results in a trajectory that curves along the outside of the road, replicating the behavior we see from human drivers.
% as we observe from naturalistic trajectories.
% Thus, projection into the \nset{} reproduces naturalistic behavior within this non-naturalistic trajectory.
Our method reproduces naturalistic behavior within this trajectory without explicitly modeling factors like comfort.

\subsection{Busy Intersection with Parked Cars}
\label{ssec:exp-recording22}
\begin{figure}[!t]
\includegraphics[width=\columnwidth]{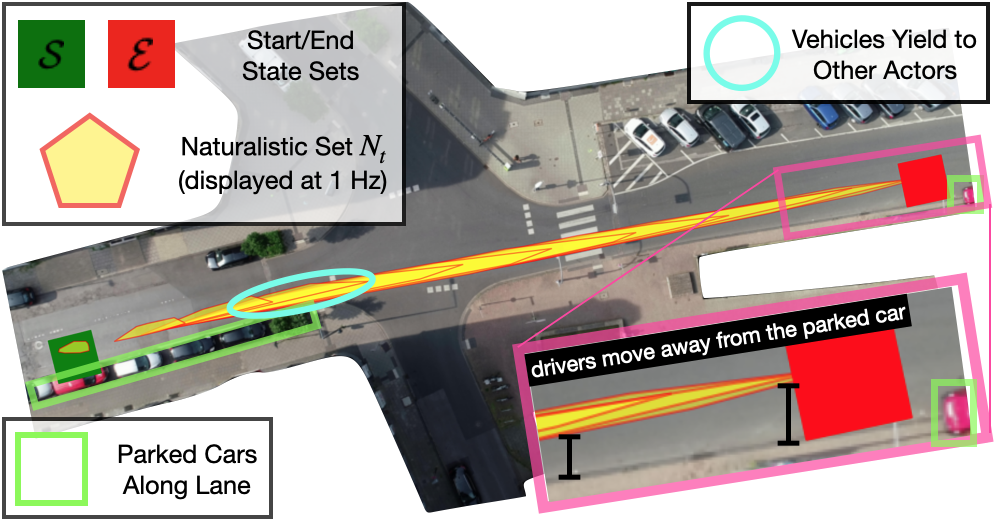}
\caption{
\label{fig:hull-r22}
% [inD, recording 22] 
We compute the \nset{} over two-dimensional position using all trajectories of moving vehicles that begin in the green square and end in the red.
We generate the \nset{} based on these trajectories and plot it.
Parked cars along the driving lane are boxed in light green.
We additionally circle (in light blue) the area of the lane in which vehicles tend to yield to crossing 
road users.
% pedestrians and bicyclists.
}
\end{figure}

Fig. \ref{fig:hull-r22} captures a main road running through a four-way intersection and a pedestrian crossing, and the road is lined with parked cars.
We define $\dataset{}$ as including all moving vehicles beginning and ending in the eastbound lane, where $\startset{}$ is given by the green square and $\finalset{}$ by the red square.
% A cross-road intersects with the main road, and there are two bike paths nearby, so many pedestrians and bikes cross the intersection.
Filtering based on these criteria results in 49 trajectories.
We expect the \nset{} to be influenced by the delays caused by naturally maintaining distance from parked cars (highlighted in light green) and by waiting for crossing pedestrians.
The \nset{} generated from $\dataset{}$ is shown in Fig. \ref{fig:hull-r22}.

\subsubsection{Analysis.}
The naturalistic sets $\{\wptset{t}\}$ begin compact but lengthen along the lane over time, indicating a variety of speeds driving along this lane.
% At the most curved point in the lane, the \nset{} covers the outside portion of the lane but not the inside.
% This observation suggests that drivers naturalistically hug the outside of a curved lane.
We make two observations about the vehicles driving along the specified path.
First, the \nset{} reflects a larger distance from parked vehicles.
% \david{I think the stronger way to make this point is to say that these polygons are longer on the side further from the parked cars, indicating that cars are traveling faster when they are on that side of the road.}
Near the beginning of the task, the \nset{} polygons are in the left portion of the lane, not centered.
More dramatically, as shown in the inset of Fig. \ref{fig:hull-r22} near the end of the task, the \nset{} is longer along the center of the lane than along the right side. This observation indicates that vehicles move faster towards the center to avoid the parked red car.
These behaviors indicate a naturalistic preference to maintain a larger distance from parked vehicles, without explicitly modeling it.

% Second, we note that the \nset{} consists of straight, long, and narrow polygons.
% As with the previous example, this reflects differences in how rapidly vehicles proceed through the intersection.
% However, in this case, the delays are caused by a need to wait for pedestrians or turning cars.
% We see the impact of this behavior in the \nset{} as the later polygons consistently stretch from before the intersection until the end of the route.

% At the beginning of the trajectory, we note that the \nset polygons are in the left portion of the lane due to the parked cars on their right.
% The \nset{} produced from $\tilde{\dataset{}}$ consists of straight, long, and narrow polygons.
% As vehicles travel faster along the lane than nay variation across it, we expect to see this shape when vehicles are driving along a lane.
% Moreover, we note that naturalistic behavior indicates a preference to maintain a larger distance from parked vehicles.
% At the beginning of the trajectory, we note that the \nset polygons are in the left portion of the lane due to the parked cars on their right.
% As they pass the intersection, this effect decreases until the end of the maneuver where they nudge over once again to avoid the parked red car near the red square.
% This behavior indicates a naturalistic preference to maintain a larger distance from parked vehicles, without us having explicitly modeled it!

Second, we note that as vehicles approach the intersection, the naturalistic polygons lengthen until they consistently stretch from before the intersection (within the blue circle) until the end of the lane.
This shape indicates that some vehicles pass through the intersection quickly, but that others wait to pass the intersection for longer.
This effect can also be observed by noting that the polygons become more opaque in this region of the lane,
% from the blue circle until the end of the lane, 
indicating that a vehicle can be anywhere in the lane at that time.
We attribute this observation to vehicles needing to wait  for other actors to cross ahead of them.
Thus, the convex hulls provide an over-approximation of the naturalistic set $\wptset{t}$ at each time. In this experiment, actors have two distinct types of behavior: either they stop and yield or they can proceed along the lane, which results in an especially large over-approximation due to the branching nature of the task.
% increasingly over-approximates the \nset{} as vehicles yield 
% \hmzh{weakly accounts for interaction}.
This over-approximation motivates further work to represent these situations with unions of polygons at each time and propose a means of projecting onto that union.
% , we note that interaction modeling may reduce the amount of over-approximation that we observe in scenarios like this one.
% \hmzh{Should I introduce a projection in this example as well?}

\section{Current Limitations \& Future Work} \label{sec:limitations}
% \topic{Next, we discuss a number of practical consideration for the usage of these method, including scenario choice and structure, non-convexity, and interactions.}

\topic{We present our approach as a practical first step towards identifying and using \nsets{}}. 
Convex hulls enable capturing naturalistic data
% enables capturing natural data 
in a computationally efficient and data efficient representation, especially when compared to learned models.
While convex hulls sometimes fail to be resilient to outliers, we note that the simplicity of the convex hull representation is extremely powerful due to the lack of assumptions on the underlying data and that  techniques exist to reject outliers when necessary \citep{fischler1981ransac}.
\subsubsection{Non-Convexity.}
\label{ssec:limitations-nonconvexity}
While our representation does not currently handle non-convexity of the naturalistic set, we note that many on-road maneuvers are not affected by this restriction (i.e. adjusting angular velocity thresholds for differently curved roads).
To describe one example where assumptions of convexity are insufficient, consider a maneuver in which vehicles drive straight along a road which has a pothole.
Naturalistic behavior may dictate that vehicles drive to either side of the pothole, meaning that no convex polygon can capture such a scenario without including behaviors that drive over the pothole.
An over-approximated set would thus be inadequate for modeling this situation.
Instead, we would need to identify two variants of naturalistic behavior within this non-convex maneuver: driving to the right of the pothole and driving to the left.
Introducing multiple naturalistic behavior subsets requires adjustments to ensure our method works effectively.
First, we would need to ensure that our method could represent each variant appropriately, and one possible method of doing so is using multiple convex hulls, though further work is needed to explore this open problem.
Second, we would need to adjust the optimization problem \eqref{eq:projection-opt} to include discrete variables for selecting between naturalistic subsets at a given time, and provide an efficient solver to the now mixed-integer problem.
% This problem becomes difficult due to the need to introduce discrete variables .
% at time steps with multiple disjoint hulls
% \david{the “non-convexity” paragraph is verbose and difficult to follow after “To model non-convexity, we”. I would give this a significant rewrite and get Adam’s advice on phrasing here.}
% To model non-convexity, we would need to identify multiple variants of the same behavior with which to construct $\dataset{}$ and then adjust the optimization problem to force a selection between multiple convex hulls at time steps with multiple disjoint hulls.

\subsubsection{Interaction.}
\label{ssec:limitations-interaction}
We note, as in Section \ref{ssec:exp-recording22}, that our technique does not currently model interaction.
We anticipate that for most cases, this case falls under the broader case of non-convexity. 
For a maneuver like that in Fig. \ref{fig:hull-r22}, which involves cars which either proceed straight unimpeded or slow down for a pedestrian crossing, we anticipate that an autonomously generated trajectory would need to select a choice closer to one of these two subsets of naturalistic behavior.
This example would lead to two possible convex hulls after a branching point in the maneuver.
In Fig. \ref{fig:hull-r22}, this branching point might in the blue circle, after which we would model two types of nonconvex naturalistic behavior as described previously.
% In this case, we can similarly use a \todo{research this and make sure I can speak intelligibly about it} branch and bound technique to project into this set.

\subsubsection{Undesirable Road Behaviors.}
\label{ssec:limitations-undesirable-behaviors}
While we note that our method is able to capture naturalistic behavior effectively, there remains an open question as to whether all naturalistic behavior should be reproduced.
For example, consider the case of an intersection in which actors regularly violate the law by ignoring stop signs.
In such a case, we would likely prefer that an autonomous planner prefer the legal requirements over enforcing the naturalistic constraints.
% As noted in Remark \ref{remark:additional-constraints}, 
As previously noted, we can encode
further preferences of this sort, such as enforcing a safety set, by introducing new constraints as in \eqref{eq:projection-opt} and \eqref{eq:projection-opt-exp}.
\section{Conclusion} \label{sec:conclusion}
In this work, we propose a method for computing a \nset{} over observations of human behavior with a set of time-indexed convex hulls.
We subsequently describe an optimization problem that, when solved, projects trajectories into the \nset{} to
produce dynamically feasible trajectories which resemble recorded human behavior.
% produce more naturalistic, dynamically consistent behavior.
We demonstrate our method on real-world naturalistic driving data, and we show that it can capture and reproduce patterns in behaviors without explicitly modeling those patterns.

\bibliography{lit_review}

\end{document}